\def\b #1{{\rm\bf #1 }}
\def\t #1{{\tilde #1 }}
\def\htd #1{{\hat{\tilde #1 }}}
\def\tdb #1{{\tilde{\rm\bf #1 }}}

\def\be{\begin{equation}}
\def\ee{\end{equation}}
\def\bea{\begin{eqnarray}}
\def\eea{\end{eqnarray}}

\def\rhoi{{\rho_{\rm i}}}
\def\rhot{{\rho_{\rm t}}}
\def\rhof{{\rho_{\rm f}}}
\def\chio{{\chi_{\rm o}}}
\def\chiI{{\chi_{1}}}
\def\chiII{{\chi_{2}}}
\def\sigmai{{\sigma_{\rm i}}}
\def\epsilono{{\epsilon_{\rm o}}}

\documentclass[12pt,pre,english,nofootinbib,showpacs,preprint,showkeys]{revtex4}

\usepackage{babel}
\usepackage{color}
\usepackage[dvips]{graphicx} 
\usepackage{amsmath,amssymb}
\begin{document}
\title{\vspace{-0.25in}
Rigorous treatment of electrostatics for spatially
varying dielectrics based on energy minimization}

\author{O. I. Obolensky}
 \altaffiliation[Also at ]{the A.F. Ioffe Institute, St. Petersburg, Russia}
\author{T. P. Doerr}
\author{R. Ray}
 \altaffiliation[Present address: ]{Department of Physics, Bose Institute,
93/1, A.P.C. Road, Kolkata 700 009, India}
\author{Yi-Kuo Yu}
 \email[Corresponding author: ]{yyu@ncbi.nlm.nih.gov}

\affiliation{
National Center for Biotechnology Information, National Library of Medicine,\\
National Institutes of Health, Bethesda, MD 20894
}

\date{\today}

\begin{abstract}
A novel energy minimization formulation of electrostatics
that allows computation of the electrostatic energy and forces to any
desired accuracy in a system with arbitrary dielectric properties
is presented. 
An integral equation for the scalar charge density is derived
from an energy functional of the polarization vector field.
This energy functional represents the true energy of the system
even in non-equilibrium states. 
Arbitrary accuracy is achieved by solving the integral equation 
for the charge density
via a series expansion in terms of the equation's kernel, 
which depends only on the geometry of the dielectrics.
The streamlined formalism operates with volume charge distributions only,
not resorting to introducing surface charges by hand. 
Therefore, it can be applied to any spatial variation of the
dielectric susceptibility, which is of particular importance
in applications to biomolecular systems.
The simplicity of application of the formalism to real problems
is shown with analytical and numerical examples.
\end{abstract}


\maketitle

\section{Introduction}

Molecular dynamics (MD) simulations of solute-solvent systems in chemistry
and biology require accurate computation of electrostatic
forces in order to obtain meaningful results. 
For practical purposes, computational efficiency is also essential,
and various formulations exist that strive to achieve a balance 
between these two requirements. The explicit solvent methods 
simulate behaviour of each single solvent molecule
which may be prohibitively expensive 
for a system of reasonable dimensions. In addition to 
having high computational costs,
explicit solvent methods are usually tailored for reproducing one of the many
physical properties of the solvent and therefore may not be well suited
for a general description of solute-solvent systems
(see \cite{Wallqvist,Tirado-Rives,Guillot} for reviews 
and performance analyses).

The alternative approach is to treat 
the solvent 
as a dielectric continuum, and the solute as a different dielectric 
object in the solvent. The dielectric properties of the solvent and the 
solute usually serve as parameters of the model. In the literature this
scheme is known as the implicit or continuum solvent method 
(for reviews see \cite{BrooksIII,Honig,Tomasi1,Cramer,Bashford}). 
Computations based on these methods are inherently faster while comparable in
accuracy with those using explicit methods, at least in the situations when
interactions between solute and solvent molecules can be neglected.
For reasons of computational efficiency,
many of the implemented implicit solvent methods make use of assumptions
which prevent improvement in accuracy even as computational resources
increase. The so-called generalized Born model is a good example
of such uncontrolled approximations (see \cite{Doerr} for a discussion).

To achieve controllable accuracy, we have recently proposed a novel 
scheme \cite{Doerr} based on determining surface charges satisfying 
the displacement field boundary condition. With this scheme, one can 
achieve any level of accuracy permitted by the available 
computing power, while remaining computationally more efficient than 
explicit solvent methods. The main idea is to treat the induced 
surface charges at the boundaries as the variables to be solved for. 
This makes the potential, expressed directly in terms of the induced 
surface charge density, continuous at the boundary. Therefore only the 
displacement field boundary condition remains, and it leads to a set 
of algebraic equations for the surface charge densities. The potential 
is obtained at no additional cost. 

One of the seeming oversimplifications in the implicit solvent methods
is the assumption of a sharp boundary between the solute and solvent.
It is known, for example, that the
solute (e.g., proteins) may strongly interact with the surrounding solvent molecules
producing the so-called hydration layer(s) \cite{Bagchi}. 
To determine electrostatic forces acting on a protein coated with such
hydration layers, one needs to find induced charges in a spatially
varying dielectric medium.
In this paper we develop a rigorous framework, based on functional minimization,
for handling spatially varying dielectrics.

Functional variation is a powerful approach 
in modern physics. 
Despite common use in quantum electrodynamics, variational techniques
in classical electrostatics are relatively rare and focus mainly
on boundary value problems for linear dielectrics \cite{Schwinger2}.
It has long been a textbook fact that the true electrostatic potential 
minimizes the system's energy for a given configuration of charges \cite{Jackson}.
A suitable energy functional can be constructed in general for any system 
of continuous media including systems with inhomogeneous and
nonlinear dielectric properties. 
For instance, free energy functionals became an important tool
in description of electrolyte solutions within the mean-field (Poisson-Boltzmann)
approach (see a recent paper \cite{McCammon} and references therein).

From our viewpoint the electrostatic potential is not 
the best choice for a minimization variable 
as it contains information about both the cause and effect, 
i.e. the source and induced charge densities. Moreover constitutive 
relations must be assumed (as in \cite{Allen}). 
And finally this approach depends on 
prior knowledge of the Green's function with boundary conditions 
suitable for the given problem. 
In contrast, we use the polarization as the fundamental function as was 
proposed by Marcus over fifty years ago \cite{Marcus}, albeit with a different functional.
 The constitutive relations are then obtained as a result of minimization
of the energy functional. The only boundary condition needed is 
that the potential goes to zero sufficiently rapidly (like inverse of the distance) 
at large distances. 

In Marcus's formulation \cite{Marcus},
the electric field and electric polarization were strongly motivated 
as the vectors defining the electric state of the system.
This formulation was aimed at the processes (charge transfer chemical reactions)
which happen on a much shorter time scale than
the molecular rearrangement in response to the changing electric field.
Marcus attempted to deal with this problem by dividing the polarization into a
fast reacting part that is proportional to the local electric field and
slowly reacting part that is not a function of the local electric field.
As a result, the free energy functional derived by Marcus contains 
several electric fields and polarizations of various 
origins.\footnote{If there is a true separation of time scales between 
various portions of the electrical response,
these excess fields should be eliminated by a proper classification
of the charges in the system:
charges that respond rapidly and whose redistribution is a function 
of the local electric field contribute to the polarization,
while charges that respond slowly are part of the so-called free charge
distribution. However, if one were allowed to combine the induced charge
due to fast-responding polarization with the frozen free charges, Marcus's 
functional becomes identical to ours.}
 
A physically sound free energy functional was proposed 
by Felderhof \cite{Felderhof} in the context of a discussion
of thermal fluctuations of the polarization and magnetization 
in dielectric magnetic media.  
However, a free energy of this type seems not to have been adopted for 
calculation of electrostatic fields until recently.
An example of numerical implementation using Felderhof's
scheme can be found
in \cite{Levy}.  There, the polarization vector field was expanded 
in a plane waves basis set. The energy functional is then an ordinary function 
of the expansion coefficients which, in turn, become the variational 
parameters of a standard multidimensional optimization problem.
Fast Fourier transforms were used to go from the real space to the reciprocal 
space representations.

An approach close to Felderhof's scheme 
was also taken in \cite{Attard} where a thermodynamic
functional was constructed with the polarization as the independent 
function. However, the techniques used there are suitable for systems 
with sharp boundaries only (the susceptibility is not considered to be a function
of coordinates, but is rather treated as a piecewise constant).

In this paper we construct an energy minimization scheme
suitable for a rigorous treatment of systems with spatially varying
dielectric functions, be they linear or nonlinear. 
In the case of linear dielectrics, our functional 
is equivalent to that proposed by Felderhof \cite{Felderhof}. 
In section \ref{funda} we
give the details of the formulation 
and describe a systematic protocol for obtaining the total charge
density. To show the versatility of the scheme we apply it
in section \ref{examples} to systems with sharp boundaries 
for which the exact solutions (or the exact equations governing 
the exact solutions) are known. 
In section \ref{numerics} we present numerical results  
for the case of two interacting
dielectric charged spheres (solutes) placed in a dielectric solvent.
We discuss the differences in force and energy between the 
situations with sharp and smooth boundaries. 
Finally we conclude
with a discussion assessing the usefulness of the method. 
Electrostatic CGS units are used throughout.

\section{Fundamental Formulation}
\label{funda}

Polarization is the response of a dielectric medium to an applied 
electric field. The phenomenon is usually visualized as the appearance 
of an induced dipole moment due to a small
shift in the relative positions of the positive and negative charge 
centers at the atomic scale \cite{Feynman}. The shift may be either 
translational or rotational or both, depending on the quantum mechanical and 
electromagnetic interactions at the atomic level. The applied electric 
fields must be weak enough not to split the atoms or 
molecules into their constituents. The system is in a state of 
equilibrium under the external electromagnetic and the intrinsic 
restoring forces. 

Quantitatively, polarization ${\b P}({\b r})$ is the 
density of induced dipole moment at location ${\b r}$. 
This density in classical electrodynamics is defined through averaging 
of dipole moments of constituent atoms/molecules
in a small volume centered around ${\b r}$.
The amount of polarization depends on the applied force and 
the susceptibility of the medium to such forces. Determination
of the susceptibility of the medium (or rather the intrinsic restoring force in the
medium) is the subject of quantum mechanics rather than classical
electrodynamics.
Polarization is thus a classical/macroscopic variable summarizing quantum mechanical
effects at the atomic/microscopic level.  
Therefore, 
we choose the polarization vector field ${\b P}({\b r})$ 
and electric field ${\b E}({\b r})$, 
in contrast to the more commonly used pair ${\b E}({\b r})$ and ${\b D}({\b r})$, 
as our fundamental variables. This choice provides a simpler connection
to the parameters determined in microscopic physics.

We express the energy as a functional $U[{\b P}]$
\be
U[{\b P}] = U_{\rm C}[{\b P}] + W[{\b P}],
\ee
\noindent where $U_{\rm C}[{\b P}]$ is the electrostatic energy of interaction
of all charges present in the system,
and $W[{\b P}]$ is the energy required to create the given
polarization vector field ${\b P}({\b r})$.

From simple considerations it can be shown \cite{Feynman,Landau}
that the variation of polarization in the vicinity of a point 
is equivalent to the presence of
an induced charge density  
$\rhoi({\b r}) = - \nabla \cdot {\b P}({\b r})$. 
Therefore, the total charge density $\rhot({\b r})$ in the medium is 
a sum of the free charge density $\rhof({\b r})$ and $\rhoi({\b r})$:
\be
\rhot(\b r)=\rhof(\b r) + \rhoi(\b r).
\ee
\noindent Then\footnote{When there is no possibility of confusion, 
we do not specify the variable for the operator $\nabla$; otherwise,
we indicate the variable by a subscript.}

\be
U_{\rm C}[{\b P}] = \frac{1}{2} 
      \int \left[\rhof({\b r}) - \nabla \cdot {\b P}({\b r})\right]
           \frac{1}{|{\b r}-{\b r}'|}
           \left[\rhof({\b r}') - \nabla \cdot {\b P}({\b r}')\right] 
      d {\b r}  d {\b r}' .
\label{UC}
\ee 

\noindent
Note first that we do not include any separate term for induced 
surface charges as was done in some of the earlier formulations of
functional minimization \cite{Marcus,Attard}. The volume charge density is the most general 
form of charge density possible. Secondly, (\ref{UC}) is the Coulomb
energy in vacuum and hence quite fundamental as opposed to the
form with the dielectric constant of the material 
in the denominator used in some of the earlier works \cite{Attard}.

The work functional $W[{\b P}]$ should contain the intrinsic self 
interaction of the polarization vector field. Here we  
consider only local contact terms for the intrinsic interactions. 
Noting that the energy functional is a scalar and assuming
${\b P} \leftrightarrow -{\b P}$ symmetry,
one can write the general work functional $W[{\b P}]$ as a 
polynomial expansion in even powers of ${\b P}$ 
(or the components $P_i$). Thus we may write,

\be
W[{\b P}] = \frac{1}{2} \int \left[
            P_i\,\left(\frac{1}{\chi ({\b r})}\right)_{ij}\,P_j
      + P_i P_j\,\left(\frac{1}{\mu({\b r})}\right)_{ijkl}\,P_k P_l
      + \cdots \right] d {\b r},
\label{WP}
\ee

\noindent
where the interaction tensors $1/\chi$, $1/\mu$, etc. describe
the linear and nonlinear dielectric properties of the media,
isotropic or anisotropic (summation over repeated indices is assumed).
The effective dielectric properties of the medium at the macroscopic 
level are now contained in these quantities. 

We emphasize that $U[{\b P}]$
is the actual energy functional unlike various other functionals 
proposed in the literature \cite{Jackson,Allen,McCammon,Briggs} which 
yield the energy or free 
energy of the system only at equilibrium. 
The equilibrium distribution 
of polarization (as well as induced charge distribution) can be
obtained by minimizing this energy functional with respect to the polarization. 
For any given external charge distribution and spatially varying 
dielectric susceptibilities one can obtain the solution analytically
or numerically. 

We may truncate 
the series in (\ref{WP}) at an order suitable for the problem at hand. 
For example, 
if the field is very weak we can retain only the quadratic term 
which corresponds to the case of linear dielectrics 
(isotropy is also assumed for the sake of simplicity of presentation):

\be
U[{\b P}] = U_C[{\b P}]
          + \frac{1}{2}\int 
            \frac{{\b P}({\b r}) \cdot {\b P}({\b r})}{\chi({\b r})} 
            d {\b r}.
\label{functional}
\ee

\noindent
Performing a functional variation with respect to
the polarization vector ${\b P}$, we arrive at
an integro-differential equation defining the equilibrium polarization
\be 
 \frac{{\b P}({\b r})}{\chi({\b r})} +   \nabla_{\b r} \int \frac{\rhof({\b r}') - \nabla \cdot {\b P}({\b r}')}
{|{\b r}-{\b r}'|} d {\b r}' = 0
\ee 
\noindent which implies 
\be 
\label{key}
{\b P}({\b r}) = \chi({\b r}) \int \left[\rhof({\b r}') - \nabla \cdot {\b P}({\b r}')\right]
 \frac{{\b r}-{\b r}'} {|{\b r}-{\b r}'|^3} d {\b r}'
= \chi({\b r}) {\b E}({\b r})\; .
\ee

\noindent
Thus the constitutive relation for a linear dielectric is obtained as a result
of functional minimization, with the expansion coefficient $\chi({\b r})$ 
turning out to be the dielectric susceptibility. 
Inserting the equilibrium polarization (\ref{key}) in (\ref{functional})
results in the well known expression for the total energy of the system:
\be
U = \frac{1}{2} 
      \int \rhof({\b r}) 
           \frac{1}{|{\b r}-{\b r}'|}
           \left[\rhof({\b r}') - \nabla \cdot {\b P}({\b r}')\right] 
      d {\b r}  d {\b r}' .
\label{energy}
\ee

Keeping two (or more) terms in the series (\ref{WP}) introduces
nonlinearity into the problem.
The energy functional in this case is given by

\be
U[{\b P}] = U_C[{\b P}]
          + \frac{1}{2}\int 
            \frac{{\b P}({\b r}) \cdot {\b P}({\b r})}{\chi({\b r})} 
            d {\b r}
          + \frac{1}{2}\int 
       \frac{\left[{\b P}({\b r}) \cdot {\b P}({\b r})\right]^2}{\mu({\b r})} 
            d {\b r}\; .
\ee

\noindent
Performing a functional variation as above we now obtain

\be \label{key1}
{\b P}({\b r}) = \chi({\b r}) {\b E}({\b r}) - 
                 2 \frac{\chi({\b r})}{\mu({\b r})} 
    \left[{\b P}({\b r}) \cdot {\b P}({\b r})\right]{\b P}({\b r})\; .
\ee

\noindent
Given that the first term on the right hand side is the dominant one, we
can obtain the solution via iteration. The first approximation would
be the same as the result for the linear dielectrics. Substituting it back
into (\ref{key1}),
we obtain at the second order of approximation,

\be
{\b P}({\b r}) = \chi({\b r}) {\b E}({\b r}) - 
                 2 \frac{\chi^4({\b r})}{\mu({\b r})} 
    \left[{\b E}({\b r}) \cdot {\b E}({\b r})\right]{\b E}({\b r})\; .
\ee

\noindent
One can continue with this to obtain a series of terms with higher and
higher powers of $[{\b E} \cdot {\b E}]$. This gives the 
desired result for nonlinear dielectrics. We should mention once more
that this solution is true for weak fields so that the higher order
terms are successively weaker. To ensure this condition we require 
$\mu({\b r}) >> \chi^3({\b r})$ to be true to any order
of approximation.

Let us now solve (\ref{key}) for the case of linear dielectrics.
We simplify the analysis by choosing the (scalar) induced density 
$\rhoi=-\nabla \cdot {\bf P}$ as our variable.

Using the relation
\be
\nabla_{\b r} \cdot \left[ {{\b r}-{\b r}' \over |{\b r}-{\b r}'|^3} \right] = 4\pi \delta({\b r} - {\b r}')
 \; ,
\ee

\noindent
we obtain from (\ref{key})

\be
\nabla \cdot \b P (\b r) = \nabla \chi(\b r) \cdot  \int {\b r - \b r' \over | \b r - \b r'|^3}
 \left[ \rhof(\b r') - \nabla \cdot \b P (\b r' ) \right] d\b r' 
+ 4\pi \chi (\b r) \left[ \rhof(\b r) - \nabla \cdot \b P (\b r ) \right] 
\ee
\noindent which implies 
\be \label{rho_i}
 \epsilon(\b r) \rhoi (\b r) = - \nabla \chi (\b r) \cdot 
 \int  \frac{{\b r}-{\b r'}} {|{\b r}-{\b r'}|^3} \left[ \rhof(\b r') + \rhoi(\b r')\right] d\b r'
 - 4\pi \chi(\b r) \rhof(\b r) \;,
\ee

\noindent
where $\epsilon = 1+ 4\pi \chi$. Equation (\ref{rho_i}) relates 
$\rhoi$ and $\rhof$. We may rewrite this equation as 

\be \label{rho_t.1}
\epsilon(\b r) \rhot(\b r) = \rhof(\b r) - 
\nabla \chi(\b r) \cdot \int \frac{{\b r}-{\b r'}} {|{\b r}-{\b r'}|^3}
 \rhot(\b r') d\b r' 
\ee
or 
\be \label{rho_t.2}
\rhot(\b r) = {\rhof(\b r) \over \epsilon (\b r)}  - 
{1\over \epsilon (\b r)}\nabla \chi(\b r) \cdot \int \frac{{\b r}-{\b r'}} {|{\b r}-{\b r'}|^3}
 \rhot(\b r') d\b r'
\ee

\noindent
This integral equation is the most general equation for total charge density in linear 
dielectric media. Note that it is a simple scalar equation for the induced charge $\rhoi$,
as opposed to (\ref{key}), a vector equation for the polarization ${\bf P}$
whose numerical solution also requires calculation of $\nabla \cdot {\bf P}$.
Once (\ref{key}) is solved for $\rhot$, the polarization field is straightforwardly
obtained by substituting $\rhot$ for $\rhof-\nabla \b P$ in (\ref{key}).
The advantages of switching to the induced charge persist even in the case of nonlinear
dielectrics.

For a system with uniform susceptibility, we obtain the expected screening
$\rhot(\b r) = {\rhof(\b r) \over \epsilon}$, so that 
$\rhoi(\b r) = -(1-\frac{1}{\epsilon})\rhof(\b r)$.
The second term in (\ref{rho_t.2}) generates induced charges due to non-uniformity
of dielectric medium. In the case of a sharp boundary, the proper limit of this term gives
rise to surface charges. A planar interface example is described in appendix \ref{appa}.

We may rewrite (\ref{rho_t.2}) in the form of an operator equation
\be \label{matrix.eq}
(\b I + \b C) \rhot = \frac{\rhof}{\epsilon},
\ee
\noindent where the operators $\b I$ and $\b C$ are defined as
\be \label{I.def}
\left[ \b I \; h \right] (\b r) =
\int \delta(\b r - \b r') h(\b r') d\b r' \;,
\ee

\be \label{C.def}
\left[ \b C \; h \right] (\b r) =
\int  {\nabla \chi (\b r) \over \epsilon(\b r)} \cdot
 {\b r -\b r' \over |\b r -\b r' |^3} h(\b r') d\b r' \;.
\ee
\noindent We will frequently make use of the kernel of this operator
defined as
\be \label{cmatrix}
\b C(\b r, \b r') = {\nabla \chi (\b r) \over \epsilon(\b r)} \cdot
 {\b r -\b r' \over |\b r -\b r' |^3}.
\ee
\noindent
Note that $\b C$ is completely determined by the geometry regardless 
of the position of the source charge.

Using the formal inversion of $\b I + \b C$
\be \label{series.sol}
\left[ \b I + \b C \right]^{-1} = \b I - \b C + \b C^2 -\b C^3 + \cdots \; ,
\ee
\noindent one may obtain the total charge density
\be \label{totalq}
\rhot = \left[ \b I - \b C + \b C^2 - \b C^3 + \cdots \right] {\rhof \over \epsilon}.
\ee
\noindent If the off-diagonal part $\b C(\b r, \b r')$ is small compared to the 
diagonal delta function, series (\ref{series.sol}) converges quickly.

\section{Three case studies \label{examples}}

In this section we apply our energy 
minization method to three examples for which the exact 
solutions or the equations governing the exact solutions are known. 

\subsection{A planar interface}
\label{plane}

Let $\chi$ depend only on one spatial variable $z$. For $z>a$, 
$\chi = \chiI$, and for $z<-a$, $\chi=\chiII$. In the 
range $-a \le z \le a$, $\chi$ is a smooth function of $z$. 
Then
\be
C(\b r, \b r') = \frac{\partial_z \chi}{\epsilon(z)} \hat z \cdot 
{\b r -\b r' \over |\b r -\b r' |^3} = 
{\partial_z \chi \over \epsilon(z)} {z - z' \over |\b r -\b r' |^3}.
\ee
\noindent Let us put a free point charge $q$ at $z = d > a$ so that
$\rhof (\b r) = q \delta (\b r - d \hat z)$. 
The total charge density (\ref{totalq}) becomes

\bea \label{totalqlong}
\rhot (\b r) 
&=& {q \over \epsilon_1 } \delta(\b r-d\hat z)  
 - {\epsilon'(z) \over 4\pi \epsilon(z)} \int {z - z' \over |\b r - \b r' |^3} 
 \delta(\b r'-d\hat z) {q \over \epsilon_1 } d\b r'  \nonumber \\
&& 
 + {\epsilon'(z) \over 4\pi \epsilon(z)} \int {z - z' \over |\b r - \b r' |^3}
   {\epsilon'(z') \over 4\pi \epsilon(z')}    {z'-z'' \over |\b r' -\b r'' |^3}
 \delta(\b r''-d\hat z) {q \over \epsilon_1 } d\b r' d\b r'' + \cdots \;,
\eea
\noindent where we have used $\epsilon=1+4\pi \chi$.

In the $a\to 0$ limit,  $\epsilon'(z) = \delta(z) (\epsilon_1-\epsilon_2)$, so
\bea
\rhot(\b r) &=& {q \over \epsilon_1 } \delta(\b r-d\hat z) - {\epsilon_1-\epsilon_2 \over 
4\pi \epsilon(z=0)} \delta(z) \int {z - z'  \over |\b r - \b r' |^3} 
 \delta(\b r'-d\hat z) {q \over \epsilon_1 } d\b r'  \nonumber \\
&& +  \left( {\epsilon_1-\epsilon_2 \over 
4\pi \epsilon(z=0)} \right)^2 \delta(z)\int {z - z' \over |\b r - \b r' |^3}
 \delta(z') {z'-z'' \over |\b r' -\b r'' |^3} \delta(\b r''-d\hat z) {q \over \epsilon_1 } d\b r' 
d\b r'' + \cdots \; .
\label{rhot.plane.series}
\eea
\noindent Note that each term from the second order on has a factor of $z \delta(z)$
which is zero for any $z$.
We finally obtain 
\be
\rhot(\b r) = {q \over \epsilon_1 } \delta(\b r-d\hat z) - {\epsilon_1-\epsilon_2 \over 
4\pi \epsilon(z=0)} \delta(z) {z-d \over |\b r- d \hat z|^3} {q \over \epsilon_1 } \;.
\label{rhot.plane.final}
\ee 
 
\noindent The surface charge density \cite{Jackson} depending on the radial
vector $\boldsymbol \rho$ in the $x-y$ plane
\be \label{planar.sol}
\sigma (\boldsymbol \rho) = {q\over 4 \pi \epsilon_1} 
\frac{2 (\epsilon_1 - \epsilon_2)}{(\epsilon_1+\epsilon_2)} 
 {d \over |\boldsymbol \rho - d \hat{z}|^3}\; ,
\ee
\noindent is then obtained by setting $\epsilon(z=0) = (\epsilon_1+\epsilon_2)/2$.
The validity of using the average dielectric constant at the boundary
is justified by the following argument. Let there be a surface charge density $\sigma$
at the boundary. It creates an electric field of magnitude $2\pi \sigma$
directed along the normal vector to the surface. Assuming that there are no
free charges at the interface, the boundary condition requires that
$(E_\perp+2\pi\sigma)\epsilon_1=(E_\perp-2\pi\sigma)\epsilon_2$, where
$E_\perp$ is a normal component of electric field produced by sources other
than $\sigma$. Therefore, $\sigma (\epsilon_1+\epsilon_2)/2=(\epsilon_2-\epsilon_1)E_\perp/4\pi$,
in agreement with setting $\epsilon(z=0) = (\epsilon_1+\epsilon_2)/2$.
In Appendix \ref{appa} we present a thorough derivation of the $a \to 0$ limit,
which arrives at the same conclusion without invoking $\delta$-functions.
It is worthwhile to point out
here that the surface charge density arises entirely from the term containing the
gradient of the susceptibility. 
Our formulation is straightforward in this respect when contrasted with 
methods that first neglect the gradient of $\chi$
and then introduce a surface charge density by hand \cite{Attard}.

\subsection{A point charge outside of a sphere}

Consider a ball of radius $a_1$ centered at the origin and 
a point charge  $q$ located at point $\b L$, $\rhof (\b r) = q \delta(\b r - \b L)$.
In this subsection, we first obtain a set of equations for the general case of spatially 
varying susceptibility,
assuming only that it changes in the radial direction. We then consider the case
of a sharp boundary and show that the simplified expressions for the induced density
coincide with the known results \cite{Menzel}.

Let the susceptibility change in the radial direction from some value $\chi_1$ inside the ball
to another value $\chio$ outside.
Gradient $\chi$ is then directed radially,
\be
\nabla \chi(\b r) = \frac{\partial \chi}{\partial r} \hat r = \frac{\epsilon'(r)}{4\pi} \hat r,
\ee 
\noindent and
we find for $C(\b r,\b r')$
\be
C(\b r,\b r') = 
\frac{\epsilon'(r)}{4\pi \epsilon(r)} \hat r \cdot {\b r - \b r' \over |\b r - \b r' |^3}=
-\frac{\epsilon'(r)}{4\pi \epsilon(r)} \partial_r {1 \over |\b r - \b r' |}.
\label{C.radial}
\ee

Let us calculate
\be
\left[\b C \cdot {\rhof \over \epsilon}\right](\b r) = 
\int d \b r' C(\b r,\b r') \frac{\rhof(\b r')}{\epsilon(\b r')}=
-\frac{q}{\epsilono}\frac{\epsilon'(r)}{4\pi \epsilon(r)} \partial_r \frac{1}{|\b r -\b L|}.
\ee

\noindent Assuming, for simplicity, that  
the point charge is located far enough from the ball,
so that $\epsilon'(r) \neq 0$ only where $r < L$ (a generalization which would lift
this condition is straightforward), 
we obtain the first order approximation for the induced charge density,
\be
\rhoi^{(1)} (\b r)\equiv \left[-\b C \cdot {\rhof \over \epsilon}\right](\b r) = 
\sum_{lm}  \rho^{(1)}_{lm} (r) Y_{lm} (\hat r) Y^*_{lm} (\hat L),
\ee
\noindent where
\be
\rho^{(1)}_{lm} (r) = \frac{4\pi}{2l + 1} \; \frac{q}{\epsilono} \frac{\epsilon'(r)}{4\pi \epsilon(r)}
\frac{l r^{l-1}}{L^{l+1}}
\label{rho.1}
\ee

\noindent and the expansion
\be
\frac{1}{|\b r_1 - \b r_2|} = \sum_{l=0}^{\infty}\sum_{m=-l}^{l} \frac{4\pi}{2l + 1} \frac{r_<^l}{r_>^{l+1}}
Y_{lm}(\hat r_<) Y^*_{lm}(\hat r_>), 
\quad r_< \equiv {\rm min}(r_1,r_2), \; r_> \equiv {\rm max}(r_1,r_2)
\label{Y.exp}
\ee
\noindent  was used. Note that any one of the spherical harmonics can bear the complex conjugation sign.

The next order is obtained by applying the operator $(-\b C)$ to $\rhoi^{(1)}$:
\be
\rhoi^{(2)} (\b r) = \left[-\b C \cdot \rhoi^{(1)} \right](\b r) =
\sum_{lm} 
\left[ \int d \b r' \left(-C(\b r,\b r')\right) \rho^{(1)}_{lm}(r')
Y_{lm} (\hat {r'})  \right] Y^*_{lm} (\hat L).
\label{rhoi.2}
\ee
\noindent The angular integration in (\ref{rhoi.2}) can be performed analytically using
(\ref{C.radial}) and (\ref{Y.exp}):
\bea
&&\int d \b r' \left(-C(\b r,\b r')\right) \rho^{(1)}_{lm}(r') 
Y_{lm} (\hat {r'}) =
\frac{\epsilon'(r)}{4\pi \epsilon(r)} \partial_r
\int d \b r' \frac{1}{|\b r - \b r'|}\rho^{(1)}_{lm} (r') 
Y_{lm}(\hat {r'}) 
\nonumber \\
&& \quad =\frac{\epsilon'(r)}{4\pi \epsilon(r)}
\sum_{l'm'} \frac{4\pi}{2l'+1} Y_{l'm'}(\hat r)
\left[ \partial_r \int_0^\infty d r'
\frac{r_<^{l'}}{r_>^{l'+1}} \rho^{(1)}_{lm}(r')
\int d \hat {r'} Y^*_{l'm'}(\hat {r'}) Y_{lm}(\hat {r'})
\right].
\eea
\noindent The orthogonality relation for the spherical harmonics,
\be
\int d \hat {r'} Y^*_{l'm'}(\hat {r'}) Y_{lm}(\hat {r'}) = \delta_{l'l}\:\delta_{m'm}
\ee
\noindent removes the sum, so we obtain
\bea
\rhoi^{(2)} (\b r) &=& 
\sum_{lm} \rho^{(2)}_{lm} (r) Y_{lm} (\hat r) Y^*_{lm} (\hat L),
\nonumber \\
\rho^{(2)}_{lm} (r)
&=&\frac{4\pi}{2l + 1} \frac{\epsilon'(r)}{4\pi\epsilon(r)}
\left[
 l \int_r^\infty \frac{r^{l-1}}{(r')^{l-1}} \rho^{(1)}_{lm}(r') d r' -
 (l+1) \int_0^r \frac{(r')^{l+2}}{r^{l+2}} \rho^{(1)}_{lm}(r') d r'
\right].
\eea

The same derivation leads us to a general recursive relation
\bea
\rhoi^{(n+1)} (\b r) &=& 
\sum_{lm}  \rho^{(n+1)}_{lm} (r) Y_{lm} (\hat r) Y^*_{lm} (\hat L),
\nonumber \\
\rho^{(n+1)}_{lm} (r)
&=&\frac{4\pi}{2l + 1} \frac{\epsilon'(r)}{4\pi\epsilon(r)}
\left[
 l \int_r^\infty \frac{r^{l-1}}{(r')^{l-1}} \rho^{(n)}_{lm}(r') d r' -
 (l+1) \int_0^r \frac{(r')^{l+2}}{r^{l+2}} \rho^{(n)}_{lm}(r') d r'
\right].
\label{rhol}
\eea
\noindent Therefore, using (\ref{totalq}), we write the induced
charge density for the general case of a sphere with a radially 
varying susceptibility  as
\bea
\rhoi (\b r) &=& 
\sum_{lm}  \rho_{lm} (r) Y_{lm} (\hat r) Y^*_{lm} (\hat L),
\nonumber \\
\rho_{lm} (r)&=& \sum_{n=1}^{\infty} \rho^{(n)}_{lm} (r),
\eea
\noindent where $\rho_{lm}^{(n)} (r)$ can be found via (\ref{rho.1}) and (\ref{rhol}).

In the limit of a sharp boundary,
\be
\epsilon'(r)=(\epsilono-\epsilon_1) \delta(r-a_1),
\ee
\noindent we immediately find that
\be
\rho^{(1)}_{lm} (r) = \frac{q}{\epsilono} \frac{\epsilono-\epsilon_1}{4\pi \epsilon(a_1)}
\frac{l a_1^{l-1}}{L^{l+1}} \delta(r-a_1),
\ee
\noindent while the higher order contributions,
\be
\rho^{(n+1)}_{lm} (r)
=\left(\frac{-1}{2}\right)^{n} \left(\frac{4\pi}{2l+1} \right)^{n+1}
\left(\frac{\epsilono-\epsilon_1}{4\pi\epsilon(a_1)}\right)^{n+1}
\frac{l a_1^{l-1}}{L^{l+1}} \delta(r-a_1),
\label{rhol.delta}
\ee
\noindent are found from (\ref{rhol}) using the generalized definition of the Dirac $\delta$-function,
\be
\int_0^\infty h(x) \delta(x) = \frac{1}{2} h(0).
\ee

Finally, we sum all the contributions to obtain the total charge density:
\bea
\rhot(\b r) &=& \left[ \left( \b I + \sum_{n=1}^\infty (- \b C )^n\right) {\rho_f\over \epsilon}\right] (\b r) \nonumber \\
 &=& {q\over \epsilono} \delta(\b r - \b L) 
 + {q\over \epsilono}
  \left({\epsilono-\epsilon_1 \over 4 \pi \epsilon(a_1) } \right) \delta(r-a_1) 
 \sum_{lm}^\infty \frac{4\pi}{2l+1} {l a_1^{l-1} \over L^{l+1}} \times \nonumber \\
&&\qquad \left[ \sum_{n=1}^\infty
  \left(- {\epsilono-\epsilon_1 \over 2 \epsilon(a_1)}{1\over 2l+1}\right)^{n-1} 
\right] \
Y_{lm} (\hat r) Y^*_{lm} (\hat L)
\label{rho.t.sum}
\eea
\noindent The sum in square brackets is a geometric series with common factor less than 1 for all $l$.
Substituting $\epsilon(a_1) = (\epsilono+\epsilon_1)/2$ again, we derive
\be
\rhot(\b r) = {q\over \epsilono} \delta(\b r - \b L) + {q\over \epsilono}
 (\epsilono-\epsilon_1) \delta(r-a_1) 
 \sum_{lm}^\infty  {l \over \left[ (l+1) \epsilono + l \epsilon_1\right]}{a_1^{l-1} \over L^{l+1}} 
Y_{lm} (\hat r) Y^*_{lm} (\hat L).
 \label{1.sphere.sol.out}
\ee

For the case in which the point charge is inside the ball, similar analysis leads to
\be
\rhot(\b r) =
{q\over \epsilon_1} \delta(\b r - \b L) + {q\over \epsilon_1}
 (\epsilono-\epsilon_1) \delta(r-a_1) 
 \sum_{lm}^\infty  {l+1 \over \left[ (l+1) \epsilono + l \epsilon_1\right]}{L^{l} \over a_1^{l+2}} 
Y_{lm} (\hat r) Y^*_{lm} (\hat L), \qquad L<a_1
 \label{1.sphere.sol.in}
\ee 

Using the addition theorem for spherical harmonics,
\be
P_l(\hat{r}\cdot \hat{L}) = \frac{4\pi}{2l+1}\sum_{m=-l}^l Y_{lm}(\hat r) Y^*_{lm}(\hat L),
\ee
\noindent and placing the point charge on the z-axis, $\b L=(0,0,L)$, 
one can further simplify the derived equations:
\be
\rhot(\b r) = {q\over \epsilono} \delta(\b r - L \hat z) + {q\over \epsilono}
 \frac{\epsilono-\epsilon_1}{4\pi} \delta(r-a_1) 
 \sum_{l}  {l (2l+1) \over \left[ (l+1) \epsilono + l \epsilon_1\right]}{a_1^{l-1} \over L^{l+1}} 
P_{l} (\cos\theta), \qquad L>a_1,
\ee
\be
\rhot(\b r) =
{q\over \epsilon_1} \delta(\b r - L \hat z) - {q\over \epsilon_1}
 \frac{\epsilono-\epsilon_1}{4\pi} \delta(r-a_1) 
 \sum_{l}  {(l+1)(2l+1) \over \left[ (l+1) \epsilono + l \epsilon_1\right]}{L^{l} \over a_1^{l+2}} 
P_{l} (\cos\theta), \qquad L<a_1.
\ee 
\noindent where $\theta$ is the polar angle of $\b r$.
These expressions provide the correct results for the surface charge
densities which can be found in \cite{Menzel}.

\subsection{Multiple charges and multiple spheres}
We now generalize to the situation of many point charges and many spheres.
In this case only the {\em exact equation}, not the exact solution, is known \cite{Doerr}.
According to the linear superposition principle, the  
induced surface charge on each sphere may be computed by using one free charge at a time
and then adding up the contributions.

Let us consider $N$ dielectric spheres of various radii and 
dielectric constants immersed inside a dielectric medium of dielectric constant $\epsilono$. 
The location of sphere $i$ is $\b R_i$, its radius is $a_i$, and its interior 
has dielectric constant $\epsilon_i$.
No two spheres are in contact with one another. 
There are $K$ point charges $q_i$ located at $\b g_i$ so that 
the free charge density reads $\rhof(\b r) = \sum_{i =1}^K q_i \delta (\b r-\b g_i )$. 
We assume that the variation of susceptibility in the vicinity of each sphere
is radial with respect to the center of that sphere:
\be
\nabla \chi(\b r) = \sum_{i=1}^N  \frac{\partial \chi}{\partial \t r_i} \htd r_i \equiv 
\sum_{i=1}^N \frac{\epsilon'(\t r_i)}{4\pi}\htd r_i \; .
\ee
\noindent Here and throughout this section we use the tilde sign to denote radius vectors
centered at the corresponding spheres, $\b r=\b R_i + \tdb r_i$.

From (\ref{rho_t.2}) we have 
\be
\rhot(\b r) = {\rhof(\b r) \over \epsilon(\b r)} - 
\sum_i \frac{\epsilon'(\t r_i)}{4\pi \epsilon(\t r_i)}
 \int \htd r_i \cdot {\b r-\b r' \over |\b r-\b r'|^3} \rho_t(\b r') d\b r'
\equiv {\rhof(\b r) \over \epsilon(\b r)} - 
\sum_i [\b C_i \rho_t](\b r)
\ee
\noindent where $\sum_i \b C_i $ plays the role of $\b C$ in (\ref{matrix.eq}).

Concentrating on the equation associated with a particular sphere $k$, 
we decompose $\rhot(\b r)$ as
\be
\rhot(\b r) = \rho_k(\b r) + {\rhof(\b r) \over \epsilon (\b r)}  + \sum_{j\ne k} \rho_j (\b r),
\ee
where 
$\rho_i(\b r)$
is the total charge density near the surface of sphere $i$. Since we consider
nonoverlapping spheres, $\b C_i \b C_j = 0$ for $i\ne j$. Therefore, when focusing on a
spatial point near sphere $k$, the only contribution to the overall charge density is 
$\rho_k (\b r)$, so $\rhot(\b r)=\rho_k(\b r)$ for $\b r$ sufficiently close to
sphere $k$.
Then in vicinity of sphere $k$ the charge density becomes
\be
\rho_k(\b r) = - \frac{\epsilon'(\t r_k)}{4\pi \epsilon(\t r_k)}
 \int \htd r_k \cdot {\b r-\b r' \over |\b r-\b r'|^3} 
 \left[\frac{\rhof(\b r')}{\epsilon(\b r')} + \rho_k (\b r') + \sum_{j\neq k} \rho_j(\b r')
 \right],
\ee
which may be expressed symbolically as 
\be
\left[ \b I + \b C_k \right] \rho_k = - \b C_k \left( {\rho_f \over \epsilon} + \sum_{j\ne k} \rho_j \right)
\ee
with 
\be
C_k(\b r, \b r' ) = \frac{\epsilon'(\t r_k)}{4\pi \epsilon(\t r_k)} 
\htd r_k \cdot {\b r-\b r' \over |\b r-\b r'|^3} \; .
\ee
This implies a symbolic solution for $\rho_k$
\be
\rho_k = -\left[ \b I - \b C_k + \b C_k^2 -\b C_k^3 + \cdots \right] \b C_k  
\left( {\rhof \over \epsilon} + \sum_{j\ne k} \rho_j \right).
\ee

Notice that the solution for the series acting on the free charges part 
will be essentially the same as that for the one sphere problem dealt 
with in the previous subsection.
Let us consider
$\b C_k\, \rho_{j\ne k}$. 

\be 
\b C_k\, \rho_j = {\epsilon'(\t r_k) \over 4\pi \epsilon(\t r_k)}
\htd r_k \cdot \int {\b r-\b r' \over |\b r-\b r'|^3} 
\rho_j(\b r') d\b r'
\ee
\noindent We switch to vectors centered on the corresponding spheres
so that the final expression is in terms of the local
 polar angle of $\tdb r_k$, which allows easier manipulation later. In this notation,
\bea
\b C_k \,\rho_j &=& {\epsilon'(\t r_k) \over 4\pi \epsilon(\t r_k)} 
  {\hat r_k} \cdot \int {\tdb r_k - (\tdb r'_j -\b L_{j \to k}) \over |\tdb r_k - (\tdb r'_j -\b L_{j \to k})|^3} 
\rho_j(\tdb r'_j) d\tdb r'_j \nonumber \\
&=& -{\epsilon'(\t r_k) \over 4\pi \epsilon(\t r_k)} \; 
 \partial_{r_k}  \!\!\int { 1 \over |\tdb r_k - (\tdb r'_j -\b L_{j \to k})|} 
\rho_j(\tdb r'_j) d\tdb r'_j
\label{m.sphere.i1}
\eea
where $\b L_{j \to k} \equiv \b R_k -\b R_j = - \b L_{k \to j}$ 
represents the vector pointing from the center of sphere $j$ to that of sphere $k$. 
Using the expansion (\ref{Y.exp}), we obtain
\be \label{m.sphere.i2}
\b C_k \,\rho_j(\b r) = - {\epsilon'(\t r_k) \over 4\pi \epsilon(\t r_k)} 
  \sum_{lm} { 4\pi l \over 2l+1} (\t r_k)^{l-1} Y_{lm}(\htd r_k) 
 \int \frac{Y_{lm}^*({\tdb r'_j -\b L_{j \to k} \over |\tdb r'_j -\b L_{j \to k}|})}{ |\tdb r'_j -\b L_{j \to k}|^{l+1} }
 \rho_j(\tdb r'_j) d \tdb r'_j \; .
\ee

The angular integral in the above equation 
was solved by Yu \cite{Yu} and employed in \cite{Doerr} where $\rho_j \propto \delta(\t r_j-a_j)$.
The process for calculating $\b C_k^n \,\rho_j$ is not affected by the detailed
result of the integration.
For now, it is sufficient to point out that the integral
 gives rise to a geometrical factor with some factorials multiplied by 
 the multipole moment $Q^j_{lm}$ of the surface charge distribution of sphere $j$. 
Denoting the integral by $\Lambda^j_{lm}(a_j,\b L_{j \to k})$,
\be
\Lambda^j_{lm}(a_j,\b L_{j \to k}) \equiv 
\int {Y_{lm}^*({\tdb r'_j -\b L_{j \to k} \over |\tdb r'_j -\b L_{j \to k}|}) 
\over |\tdb r'_j -\b L_{j \to k}|^{l+1} }
 \rho_j(\tdb r'_j) d \tdb r'_j \; ,
\ee
we may then write 
\be
\b C_k \,\rho_j(\b r) = - {\epsilon'(\t r_k) \over 4\pi \epsilon(\t r_k)} \sum_{lm}  {4\pi \over 2l+1} l  
\Lambda^j_{lm}(a_j,\b L_{j \to k})
 \t r_k^{l-1} Y_{lm}(\htd r_k) \;.
\ee
For the case of sharp boundaries between the spheres and the external
medium, one then obtains
\be \label{m.sphere.t1}
\b C_k\, \rho_j(\b r) = - {\epsilono-\epsilon_k \over 4\pi \epsilon(a_k)}\delta(\t r_k-a_k) 
  \sum_{lm}  {4\pi \over 2l+1}\left[l a_k^{l-1} \Lambda^j_{lm}(a_j,\b L_{j \to k})\right]
 Y_{lm}(\htd r_k) \;.
\ee

Applying the $\b C_k$ operator once again and performing the integration
in the radial direction, we find
\be
\b C_k^2 \,\rho_j (\b r) =  - \left({\epsilono-\epsilon_k \over 4\pi \epsilon(a_k)}\right)^2 \!\!
 {\delta(\t r_k-a_k)\over 2} 
 \int {d\htd r'_k \over |2-2\htd r_k \cdot \htd r'_k|^{1/2}}   
\sum_{lm} {4\pi \over 2l+1}\left[l a_k^{l-1} \Lambda^j_{lm}(a_j,\b L_{j \to k})\right] Y_{lm}(\htd r'_k)
\ee
After performing the angular integration, 
$\b C_k^2 \,\rho_j (\b r)$ becomes
\be
\b C_k^2 \,\rho_j (\b r) = - \left({\epsilono-\epsilon_k \over 4\pi \epsilon(a_k)}\right)
 \delta(\t r_k-a_k)    
\sum_{lm} \left({\epsilono-\epsilon_k \over 2 \epsilon(a_k) (2l+1)}\right) {4\pi \over 2l+1}
\left[l a_k^{l-1} \Lambda^j_{lm}(a_j,\b L_{j \to k})\right] Y_{lm}(\htd r_k)
\label{m.sphere.t2}  
\ee
It is easy to see that this process continues and one ends up having 
\be \label{m.sphere.tn} 
\b C_k^n \,\rho_j (\b r) =  - \left({\epsilono-\epsilon_k \over 4\pi \epsilon(a_k)}\right) \!\!
 \delta(\t r_k-a_k)    
\sum_{lm} \left({\epsilono-\epsilon_k \over 2 \epsilon(a_k) (2l+1)}\right)^{n-1}
 \!\!\! \!\!\!{4\pi \over 2l+1}
\left[l a_k^{l-1} \Lambda^j_{lm}(a_j,\b L_{j \to k})\right] Y_{lm}(\htd r_k)
\ee
and therefore
\bea
\sum_{n=1}^\infty (-\b C_k)^n\rho_j &=& 
 - \left({\epsilono-\epsilon_k \over 4\pi \epsilon(a_k)}\right) 
 \delta(\t r_k-a_k)  \times \nonumber \\
&& \times   
\sum_{lm} \left[ \sum_{n=1}^\infty (-1)^n \! \left({\epsilono-\epsilon_k \over 2 \epsilon(a_k) (2l+1)}\right)^{n-1}
 \! \right]
  \!\! {4\pi \over 2l+1}
\left[l a_k^{l-1} \Lambda^j_{lm}(a_j,\b L_{j \to k})\right] Y_{lm}(\htd r_k) \nonumber \\
&=& - \left({\epsilono-\epsilon_k \over 4\pi }\right)
 \delta(\t r_k-a_k)  \times \nonumber \\
&& \times   
\sum_{lm} \left[ {(2l+1) \over (l+1) \epsilono + l \epsilon_k }
 \right]
 {4\pi \over 2l+1}
\left[l a_k^{l-1} \Lambda^j_{lm}(a_j,\b L_{j \to k})\right] Y_{lm}(\htd r_k) \; ,
 \label{m.sphere.final}
\eea
where $\epsilon(a_k) = (\epsilono+ \epsilon_k)/2$ is used.
We are now in a position to write down the full solution using 
(\ref{1.sphere.sol.out}), (\ref{1.sphere.sol.in}), and (\ref{m.sphere.final}).
Defining ${\cal I}_k\equiv \left\{ i \left| a_k>|\b g_i-\b R_k|\right.\right\}$
and ${\cal O}_k\equiv \left\{ i \left| a_k<|\b g_i-\b R_k|\right.\right\}$
to be the sets of charges inside and outside sphere $k$, respectively, we find
\bea
\rho_k(\tdb r_k) & = & - \sum_{{\cal I}_k} {q_i \over \epsilon_k}
 \left(\epsilono-\epsilon_k \right) \delta(\t r_k-a_k) 
 \sum_{lm} {(l+1) \over \left[ (l+1) \epsilono + l \epsilon_k\right]}{|\b g_i-\b R_k|^{l} \over a_k^{l+2}}  
Y^*_{lm}\left(\frac{\b g_i-\b R_k}{|\b g_i-\b R_k|}\right)
Y_{lm}(\htd r_k) \nonumber \\ 
& + & \sum_{{\cal O}_k}
{q_i \over \epsilon(\b g_i)}
 \left(\epsilono-\epsilon_k \right) \delta(\t r_k-a_k) 
 \sum_{lm} {l \over \left[ (l+1) \epsilono + l \epsilon_k\right]}{ a_k^{l-1} \over |\b g_i - \b R_k|^{l+1}}  
Y^*_{lm}\left(\frac{\b g_i-\b R_k}{|\b g_i-\b R_k|}\right)
Y_{lm}(\htd r_k) \nonumber \\
& - & \sum_{j \neq k} \left({\epsilono-\epsilon_k \over 4\pi }\right)
 \delta(\t r_k-a_k)  
\sum_{lm} \left[ {(2l+1) \over (l+1) \epsilono + l \epsilon_k }
 \right]
 {4\pi \over 2l+1}
\left[l a_k^{l-1} \Lambda^j_{lm}(a_j,\b L_{j \to k})\right] Y_{lm}(\htd r_k) 
\label{final.rho.k}
\eea
which, with appropriate rotations and taking a single point charge at
the center of each sphere,
is equivalent to (11) in \cite{Doerr}.

\section{Numerical case study}
\label{numerics}

In this section we present results of numerical computations 
comparing the force
between two charged identical spheres with sharp boundaries to
the force between two charged identical spheres with smeared boundaries.
For brevity, the spheres with smeared boundaries will be called ``fuzzy spheres"
and the spheres with sharp boundaries will be called ``rigid spheres".
The dielectric constant $\epsilon_1 = 4$ inside the spheres and $\epsilono = 80 $ outside.
For the fuzzy spheres there is an interface region $r_0 -\delta r < r< r_0 +\delta r$ 
in which the dielectric constant changes smoothly from $\epsilon_1$ to $\epsilono$ 
in the radial direction (with respect to the center of the corresponding sphere).

\begin{figure}
\includegraphics{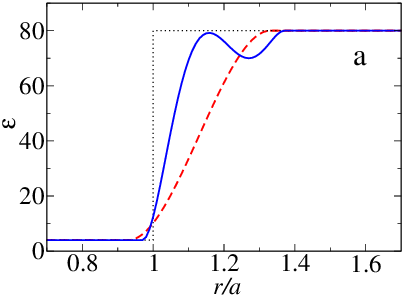}
\hspace{1cm}
\includegraphics{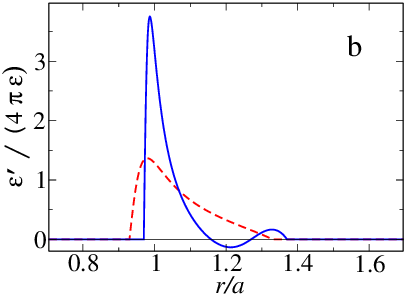}
\caption{\label{fig:epsilon} 
Radial dependence of (a) the dielectric constant $\epsilon(r/a)$ and (b) 
$\epsilon'(r/a)/(4\pi\epsilon(r/a))$
for a monotonic step
(red broken line, Eq.~(\ref{epsilonCubic})) and 
for a non-monotonic step simulating a hydration layer (blue solid line). 
The dielectric constant changes smoothly from $\epsilon_1 = 4$ inside
the sphere to $\epsilono = 80 $ outside. 
The effective radii $r_0 = 1.13a$ and $r_0 = 1.17a$, respectively,
are chosen so that the Born solvation energy in each case is 
equal to that in the case of
a sharp boundary at radius $a$ (shown with dotted line).
The half-width of the steps $\delta r = 0.2 a$.} 
\end{figure}

The simplest polynomial smoothly connecting $\epsilon_1$ and $\epsilono$, i.e.,
satisfying the conditions $\epsilon(r_0-\delta r) = \epsilon_1$, $\epsilon(r_0+\delta r) = \epsilono$,
$\epsilon'(r_0-\delta r) = \epsilon'(r_0+\delta r) = 0$, is cubic,
so that the dielectric constant can be defined around each sphere as

\bea
\epsilon(r)&=& \epsilon_1, \qquad r<r_0 -\delta r \nonumber \\
\epsilon(r)&=&
\left[\frac{\left(r-r_0\right)^3}{\delta r^3} -3\frac{r-r_0}{\delta r}
\right] \frac{\epsilon_1-\epsilono}{4} + \frac{\epsilon_1+\epsilono}{2}, 
\qquad r_0-\delta r \le r \le r_0+\delta r \nonumber \\
\epsilon(r)&=& \epsilono, \qquad r>r_0 +\delta r.
\label{epsilonCubic}
\eea
\noindent With a fifth order polynomial one can request additionally that
$\epsilon(r_0 + \delta r_{\rm H}) = \epsilon_{\rm H}$ and 
$\epsilon'(r_0 + \delta r_{\rm H}) = 0$. Letting $\epsilon_{\rm H} = 70$
and $\delta r_{\rm H} = 0.5 \delta r$ yields a non-monotonic profile, which may be used
to simulate the hydration layer phenomenon in bio-macromolecules 
and clusters (see Fig.~\ref{fig:epsilon}).

Let there be point charges $q_1$ and $q_2$ at the centers of spheres 1 and 2, respectively.
The induced charge density is found for rigid spheres as the self-consistent solution
of (\ref{final.rho.k}) for $\rho_1(\t r_1)$ and $\rho_2(\t r_2)$.
Of course, (\ref{final.rho.k}) simplifies dramatically in the case of two spheres
and two free charges. For fuzzy spheres, one has to use a continuous version of (\ref{final.rho.k})
in which summation over $n$ in (\ref{m.sphere.final}) is carried out numerically with
the $n^{\rm th}$-order terms (\ref{m.sphere.tn}) calculated recursively via numerical
integration, analogously to the method for a point charge outside a sphere,
see (\ref{rhol}), (\ref{rhol.delta}) and (\ref{rho.t.sum}).
Notice that the $l=0$ components of the induced densities can only be produced
by the free charge inside the corresponding sphere. Notice also that the free charges
in the centers of the spheres induce only $l=0$, i.e. spherically-symmetric, components. 
For these reasons it is convenient to distinguish the $l=0$ and $l \ne 0$ components
of the induced charge density.

In accordance with (\ref{energy}), the total energy of the system consists
of the following terms: 

(i) interaction between the point charges (screened by $\epsilon_1$), 

\be
\frac{q_1 q_2}{\epsilon_1 L},
\label{term.i}
\ee
\noindent where $L$ is the length of the vector $\b L_{1\to 2} = - \b L_{2\to 1}$,
connecting the centers of the two spheres,

(iia) interaction between each point charge
and the $l=0$ component of the induced charge in the interface region of the other sphere, 
\be
\frac{1}{2}\left( 
q_1 \int \frac{\left.\rho_2(\tdb r_2)\right|_{l=0}}{|\tdb r_2+\b L_{1\to2}|} \, d \tdb r_2
+q_2 \int \frac{\left.\rho_1(\tdb r_1)\right|_{l=0}}{|\tdb r_1+\b L_{2\to1}|} \, d \tdb r_1
\right),
\label{term.iia}
\ee

(iib) interaction between each point charge
and the $l\ne0$ components of the induced charge in the interface region of the other sphere, 
\be
\frac{1}{2}\left( 
q_1\int \frac{\left.\rho_2(\tdb r_2)\right|_{l\ne0}}{|\tdb r_2+\b L_{1\to2}|} \, d \tdb r_2
+q_2\int \frac{\left.\rho_1(\tdb r_1)\right|_{l\ne0}}{|\tdb r_1+\b L_{2\to1}|} \, d \tdb r_1
\right),
\label{term.iib}
\ee

(iiia) interaction between each point charge
and the $l=0$ component of the induced charge in the interface region of the same sphere, 
\be
\frac{1}{2}\left( 
q_1 \int \frac{\left.\rho_1(\tdb r_1)\right|_{l=0}}{\t r_1} \, d \tdb r_1
+q_2 \int \frac{\left.\rho_2(\tdb r_2)\right|_{l=0}}{\t r_2} \, d \tdb r_2
\right),
\label{term.iiia}
\ee

(iiib) interaction between each point charge
and the $l\ne0$ components of the induced charge in the interface region of the same sphere,
\be
\frac{1}{2}\left( 
q_1\int \frac{\left.\rho_1(\tdb r_1)\right|_{l\ne0}}{\t r_1} \, d \tdb r_1
+q_2 \int \frac{\left.\rho_2(\tdb r_2)\right|_{l\ne0}}{\t r_2} \, d \tdb r_2
\right),
\label{term.iiib}
\ee

The sum of terms 
(i) and (iia)
is equal to the energy of interaction of two point charges
in dielectric medium $\epsilono$ 
\be
\frac{q_1 q_2}{\epsilono L}.
\label{term.i.iia}
\ee
\noindent
This energy is the same for rigid and fuzzy spheres.
In contrast, terms (iib) are different for rigid and fuzzy spheres and are the main source
of differences in the forces in these two situations.
Finally, terms (iiib) are zero for the point charges located at the centers of the spheres, while
terms (iiia) are the Born solvation energy in this case.

\begin{figure}
\vspace{1cm}
\includegraphics{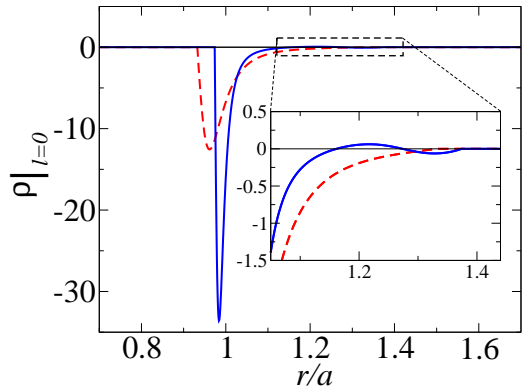}
\caption{\label{fig:rho} 
Radial dependence of  
the induced electric density
$\left. \rho(r/a)\right|_{l=0}$ for the monotonic (red broken line) 
and non-monotonic (blue solid line) steps shown in Fig.~\protect{\ref{fig:epsilon}}.
The density is normalized by the value of the point charge in the center of the sphere.
The inset magnifies a small, oscillatory feature associated with the non-monotonic step.}
\end{figure}

Born solvation energies are quite different for rigid and fuzzy spheres, since for fuzzy spheres the
induced charge density tends to accumulate near the inner boundary of the interface region. 
Indeed, the operator $\b C$ is proportional to $\epsilon'(r)/\epsilon(r)$ and 
$\epsilon(r_0-\delta r) = \epsilon_1 \ll \epsilon(r_0+\delta r) = \epsilono$. This asymmetry
is present at each order $n$ and is preserved after the summation over $n$.
Radial dependences of the $l=0$ components of the induced densities are illustrated 
in Fig.~\ref{fig:rho}.
On the other hand, fuzzy and rigid spheres model the same physical objects, so it is
reasonable to assume that whatever profile of the dielectric constant is chosen, the
Born solvation energy should remain the same. For this reason, we adjust 
the effective radius $r_0$ for each profile of the dielectric constant so that the Born
solvation energy is equal to that of a rigid sphere of unit radius, see Fig.~\ref{fig:epsilon}. 

\begin{figure}
\vspace{1cm}
\includegraphics{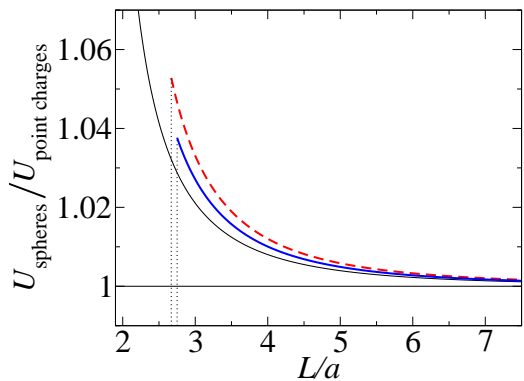}
\caption{\label{fig:energies} 
Energy of interaction between two spheres with sharp (thin line) and smeared (thick lines)
boundaries.  The red broken thick line corresponds to the case of the monotonic radial dependence
of the dielectric constant, while the blue solid thick line corresponds to the non-monotonic 
radial dependence shown in Fig.~\protect{\ref{fig:epsilon}}. Free charges of the same sign
are located at the centers of the spheres.
The energies are normalized by the Coulomb energy of these point charges
in the uniform dielectric medium $\epsilono$. The vertical dotted lines indicate
the contact points.}
\end{figure}

\begin{figure}[tb]
\vspace{1cm}
\includegraphics{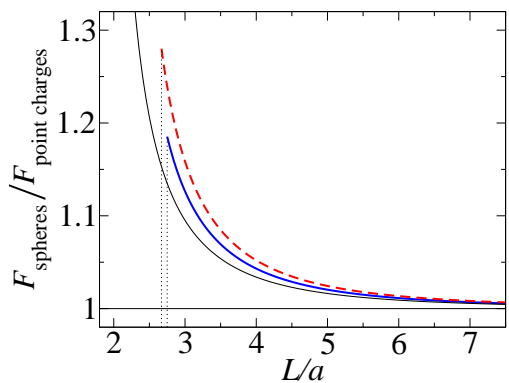}
\caption{\label{fig:forces} 
Interaction forces between two spheres with sharp and smeared
boundaries.  
The line identifications are same as in Fig.~\protect{\ref{fig:energies}}.
}
\end{figure}

In Fig.~\ref{fig:energies} we present the dependence of the interaction energy on distance
for a pair of rigid spheres and for two pairs of fuzzy spheres, with
monotonic and non-monotonic behaviour of the dielectric function in the interface region,
respectively.
The energies are normalized to the energy of interaction of point charges (\ref{term.i.iia}).
The forces between two fuzzy spheres and between two rigid spheres
are shown in Fig.~\ref{fig:forces}. The forces are normalized by the interaction 
force between two point charges.  We note that the seemingly weaker effect for the
fuzzy spheres with non-monotonic $\epsilon(r)$ dependence is due to the fact that
$\epsilon(r)$ changes faster near the inner surface of the interface region 
to make room for the feature representing the hydration layer. This makes the
fuzzy spheres with non-monotonic $\epsilon(r)$ dependence effectively more similar
to rigid spheres for fixed $\delta r$ (compare the charge density distributions in Fig.~\ref{fig:rho}).

\begin{figure}[tb]
\vspace{1cm}
\includegraphics{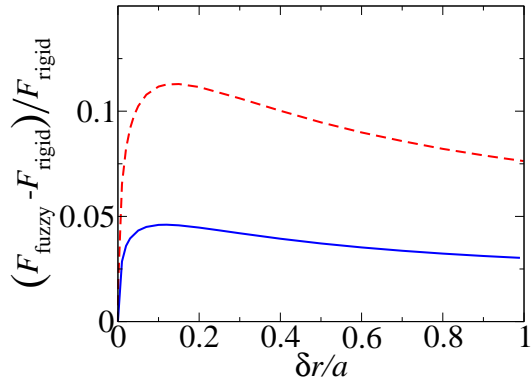}
\caption{\label{fig:effect} 
Maximum difference in interaction forces between two spheres with smeared and two spheres with sharp
boundaries, occuring at the contact point $2(r_0+\delta r)$, 
as a function of half width of the interface region $\delta r$.
The forces are normalized by the interaction force 
between two spheres with sharp boundaries.  
The red broken line corresponds to the case of the monotonic radial dependence
of the dielectric constant, while the blue solid line corresponds to the non-monotonic 
radial dependence shown in Fig.~\protect{\ref{fig:epsilon}}. 
}
\end{figure}

For very thin interface regions ($\delta r \to 0$), the forces between two rigid
and two fuzzy spheres are equal, as expected. For fuzzy spheres with moderate interface
region widths, the repulsion increases with the width.
However, this trend quickly saturates (Fig.~\ref{fig:effect}). 
Qualitatively, this saturation can be explained
by two opposing effects. 
The increase in the interface width increases
the size of the spheres thereby strengthening the repulsion. On the other hand,
the induced charge density tends to concentrate 
near the inner surface of the interface which remains around $r=a$ 
to maintain constant Born solvation energy. Therefore, the 
bulk of the induced charge on one sphere becomes farther from
that of the other sphere,
hence weakening the repulsion.



We finally note, that if the point charges are located away from the centers of the spheres,
the terms (iiib) depend on the relative position and orientation
of the spheres. In this case one can still define the Born solvation energies
as the sum of terms (iiia) and (iiib) at large separations,
but the terms (iiib) would contribute to the difference of interaction forces/energies
between the rigid and fuzzy spheres.

\section{Conclusions}
\label{disc}

We have presented an energy minimization formulation of electrostatics
that allows computation of the electrostatic energy and forces to any
desired accuracy in a system with arbitrary dielectric properties. 
We have derived an integral equation for the scalar charge density 
from an energy functional of the polarization vector field.
This energy functional represents the true energy of the system
even in non-equilibrium states. 
Arbitrary accuracy is achieved by solving the integral equation for the charge density
via a series expansion in terms of the equation's kernel, 
which depends only on the geometry of the dielectrics.
The streamlined formalism operates with volume charge distributions only,
not resorting to introducing surface charges by hand as is done in
various other studies of electrostatics via energy minimization.
Therefore, it can be applied to arbitrary spatial variations of the
dielectric susceptibility.
The simplicity of application of the formalism to real problems
has been shown with three analytic examples and with a numerical case study.
We found that finite boundary widths introduce a measurable correction
to the interaction forces as compared to sharp boundary case.
For two charged identical spheres the correction is about 10\%.

The formalism has various potential applications in modeling 
electrostatic interactions between solvated molecules:
it enables one to go beyond the widely used simplification 
of atoms and molecules as dielectric balls immersed in a dielectric 
solvent, as was first suggested by Born in the early twenties of the last century \cite{Born}.
For example, the description of an aqueous solvent as a continuous and 
homogeneous dielectric medium fails to account for the strong dielectric 
response of water molecules around charges. 
Normally, charged ions and surfaces give rise to hydration layers
by orienting and displacing surrounding water molecules.
These hydration phenomena are very important in many biological processes 
such as protein folding, protein crystallization, and interactions between 
charged biopolymers inside the cell. With our formalism one can now consider 
arbitrary structures for such hydration layers and arrive at
a possibly more realistic and reliable analysis of the molecular mechanisms
in bio-chemical interactions. 

Applied to MD simulations, this formulation is still an implicit solvent scheme, 
and the position-dependent susceptibility is therefore a model parameter (indeed, the only one).  
To obtain an estimate of the macroscopic dielectric 
susceptibility at the molecular level or at the intermolecular 
boundaries one has to explore physics at the atomic level and introduce some coarse graining. 
Given that the dielectric susceptibility is related to the charge 
fluctuations as a response to external perturbations, one can 
estimate susceptibilites through the study of 
linear/nonlinear response. For example, the dielectric susceptibility 
can be related to the correlations of the net system
dipole moment and local polarization density \cite{Stern}. 
A fully quantum mechanical treatment of solvation of biological systems 
might be hindered by limits of numerical accuracy \cite{Kohn.Nobel}
and will demand much more computational power than currently available.
We believe that quantum
mechanics, in particular, density functional theory, can in principle be used
to calculate the local dielectric susceptibility which in turn should
be used as input for the implicit solvent methods, such as the one described
in this paper.

\section*{Acknowledgements}
This research was supported by the Intramural Research Program of the NIH, NLM. 
The computations were performed on the Biowulf Linux cluster 
at the National Institutes of Health, Bethesda, MD (http://biowulf.nih.gov).

\appendix

\section{Sharp boundary limit in the planar interface problem} \label{appa}

Let us demonstrate how a rigorous limiting procedure applied to (\ref{rhot.plane.series})
produces correct expression for the surface charge density in the case
of sharp planar interface.
The surface charge is found by integrating the charge density over the range $-a \le z \le a$
in which $\chi$ changes from $\chiII$ to $\chiI$, 
and then taking the limit $a \rightarrow 0$.

We return to (\ref{totalqlong}) and, making use of 
the azimuthal symmetry of the problem, expand the kernels
in terms of Bessel functions $J_m$ \cite{Jackson}, 

\bea
\label{kernel1}
\frac{1}{|\b r - \b r' |} &=& \sum_{m=-\infty}^{\infty} \int_0^{\infty}
    e^{i\,m (\phi - \phi')}\,J_m(k \rho) \, J_m(k \rho') 
    e^{-k(z_{>} - z_{<})} d k \\
\label{kernel2}
\frac{1}{|\b r - d\,\hat{z} |} &=& \int_0^{\infty} J_0(k \rho) e^{-k(d-z)} d k .
\eea

\noindent
Here the position vectors $\b r$ and $\b r'$ are represented via
the polar vectors $\boldsymbol\rho$ and $\boldsymbol\rho'$ in the $z=0$ plane,
$\b r = \boldsymbol\rho + z \hat z$ and $\b r' = \boldsymbol\rho' + z' \hat z$.
The polar vectors are in turn defined through their lengths
$\rho = \sqrt{x^2+y^2}$ and $\rho' = \sqrt{x'^2+y'^2}$ and
their azimuthal angles $\phi$ and $\phi'$.
The notation $z_>$ ($z_<$) is used for the greater (lesser) of the corresponding $z$ and $z'$.

We now treat each of the terms in the expansion of (\ref{totalqlong})
separately. The first term is the screened point charge. All other terms
form the induced charge density at the interfacial region.
The first contribution to the induced charge density is given by

\be \label{rho1}
\rhoi^{(1)} (\b r) = -\frac{q}{\epsilon_1} 
   \frac{\epsilon'(z)}{4 \pi \epsilon(z)} \frac{z - d}{|\b r - d\,\hat{z} |^3}.
\ee

\noindent The corresponding surface charge density is 

\be 
\sigmai^{(1)} (\boldsymbol \rho) = -\frac{q}{\epsilon_1} 
\lim_{a \rightarrow 0} \int_{-a}^a 
\frac{\epsilon'(z)}{4 \pi \epsilon(z)}
\left[ \frac{ - d}{|\boldsymbol \rho - d\,\hat{z} |^3} + {\cal O}(z)
\right] d z.
\ee

\noindent All the ${\cal O}(z)$ terms vanish since for any bounded function $h(z)$

\be
\lim_{a \rightarrow 0} \int_{-a}^a z^n h(z) d z \leq 
\lim_{a \rightarrow 0} a^n \int_{-a}^a |h(z)| d z = 0, \;\;\;\; \forall \;\; n>0.
\ee

Thus,
\be \label{sigma1}
\sigmai^{(1)} (\boldsymbol \rho) = 
\frac{q}{4 \pi \epsilon_1} \frac{d}{|\boldsymbol \rho - d\,\hat{z} |^3} (f_1 - f_2).
\ee 
\noindent
Here we have used the notations $f(z) = \ln [\epsilon(z)]$, 
$f_1 = f(a) = \ln [\epsilon_1]$ and $f_2 = f(-a) = \ln [\epsilon_2]$.

We can similarly evaluate all the other contributions to the induced surface
charge density. The second contribution to the induced charge density is

\be
\rhoi^{(2)} (\b r) = \frac{q}{\epsilon_1} 
 \frac{\epsilon'(z)}{4 \pi \epsilon(z)} \int
 \frac{z - z'}{|\b r - \b r' |^3}
 \frac{\epsilon'(z')}{4 \pi \epsilon(z')} \frac{z' - d}{|\b r' - d\,\hat{z} |^3}
 \rho' d \rho' d \phi' d z' 
\ee 
\noindent Using (\ref{kernel1}), (\ref{kernel2}), and the completeness relation 
for Bessel functions \cite{Jackson},

\be \label{orthoJ}
\int_0^{\infty} J_m (k \rho) J_m (k' \rho) \rho d \rho = \frac{1}{k} \delta (k - k'),
\ee
\noindent we obtain, after integration over $\phi'$ and $\rho'$,

\bea
\rhoi^{(2)} (\b r) &=& \frac{q}{\epsilon_1}
 \frac{\epsilon'(z)}{4 \pi \epsilon(z)} \frac{d}{d z} \int
 \frac{\epsilon'(z')}{4 \pi \epsilon(z')} 
\int_0^{\infty} J_0 (k \rho) e^{-k (d-z')} e^{-k(z_> - z_<)} 2 \pi d k d z'
\nonumber \\
 &=& \frac{q}{\epsilon_1} 
 \frac{\epsilon'(z)}{4 \pi \epsilon(z)} \frac{1}{2} 
\int_0^{\infty} k d k e^{-k (d-z)} J_0 (k \rho) \times 
\nonumber \\
 && \hskip 1 in
\left[ \int_z^a \frac{\epsilon'(z')}{\epsilon(z')} d z'   
 -\int_{-a}^z \frac{\epsilon'(z')}{\epsilon(z')} e^{-2k(z-z')} d z' 
\right].
\eea

\noindent The corresponding surface charge density is then

\bea \label{sigma2}
\sigmai^{(2)} &=& \frac{q}{\epsilon_1} 
\lim_{a \rightarrow 0} \int_{-a}^a dz 
 \frac{\epsilon'(z)}{4 \pi \epsilon(z)} \frac{1}{2} 
\int_0^{\infty} k d k e^{-k (d-z)} J_0 (k \rho) \times 
\nonumber \\
 && \hskip 1 in
\left[ \int_z^a \frac{\epsilon'(z')}{\epsilon(z')} d z'   
 -\int_{-a}^z \frac{\epsilon'(z')}{\epsilon(z')} e^{-2k(z-z')} d z' 
\right].
\eea
\noindent Applying to (\ref{sigma2}) the same argument used in deriving (\ref{sigma1}),
\be 
\sigmai^{(2)} = \frac{q}{\epsilon_1} 
\int_0^{\infty} k d k e^{-k d} J_0 (k \rho)
\lim_{a \rightarrow 0} 
\int_{-a}^a dz \frac{\epsilon'(z)}{4 \pi \epsilon(z)} 
\frac{1}{2} 
\left[ \int_z^a \frac{\epsilon'(z')}{\epsilon(z')} d z'   
 -\int_{-a}^z \frac{\epsilon'(z')}{\epsilon(z')} d z' 
\right].
\ee

\noindent The integral over $k$ is evaluated using (\ref{kernel2}) as
\be \label{kernel3}
\int k J_0 (k \rho) e^{-k d} d k
= \left. \frac{d}{dz} \int J_0 (k \rho) e^{-k (d-z)} d k \right|_{z=0}=
\left. \frac{d}{dz} \frac{1}{|\b r - d\,\hat{z} |} \right|_{z=0}= 
\frac{d}{|\boldsymbol\rho - d\,\hat{z} |^3}.
\ee

\noindent Then
\be 
\sigmai^{(2)} = \frac{q}{\epsilon_1} 
\frac{d}{|\boldsymbol\rho - d\,\hat{z} |^3}
\lim_{a \rightarrow 0} 
\int_{-a}^a dz \frac{\epsilon'(z)}{4 \pi \epsilon(z)} 
\left[ \frac{1}{2} (f_1+f_2) - f(z) \right].
\ee

\noindent Finally, we obtain that $\sigmai^{(2)} = 0$,
\be
\sigmai^{(2)} = \frac{q}{4\pi \epsilon_1} 
\frac{d}{|\boldsymbol\rho - d\,\hat{z} |^3}
\int_{f_2}^{f_1} df \left[ \frac{1}{2} (f_1+f_2) - f(z) \right] = 0.
\ee

Analogously, the expressions for the induced surface charge densities 
up to the fifth order are found to be

\bea
\sigmai^{(1)} &=& \frac{q}{4 \pi \epsilon_1}
 \frac{d}{|\boldsymbol \rho  - d\,\hat{z} |^3} 
 (f_1 - f_2), \nonumber \\
\sigmai^{(2)} &=& 0,
  \nonumber \\
\sigmai^{(3)} &=& \frac{q}{4 \pi \epsilon_1}
 \frac{d}{|\boldsymbol \rho - d\,\hat{z} |^3} 
 \frac{-1}{12}(f_1 - f_2)^3, \nonumber \\
\sigmai^{(4)} &=& 0,
  \nonumber \\
\label{sigmans}
\sigmai^{(5)} &=& \frac{q}{4 \pi \epsilon_1}
 \frac{d}{|\boldsymbol \rho - d\,\hat{z} |^3} 
 \frac{1}{120}(f_1 - f_2)^5.
\eea

In general, the surface charge density is of the form
\bea
\label{sigman}
\sigmai^{(n)}(z) &=&-\frac{q}{\epsilon_1} \lim_{a \rightarrow 0} \int_{-a}^{a} dz
   \frac{\epsilon'(z)}{4 \pi \epsilon(z)} \frac{z - d}{|\b r - d\,\hat{z} |^3} \times
\nonumber \\
   && \qquad \frac{1}{2} \left[\int_z^a \frac{\epsilon'(z')}{\epsilon(z')} g^{(n-1)}(f(z')) d z' 
       - \int_{-a}^z \frac{\epsilon'(z')}{\epsilon(z')} g^{(n-1)}(f(z')) d z'
   \right]
\nonumber \\
             &=&-\frac{q}{\epsilon_1} \lim_{a \rightarrow 0} \frac{1}{2} \int_{-a}^{a} dz
   \frac{\epsilon'(z)}{4 \pi \epsilon(z)} \frac{z - d}{|\b r - d\,\hat{z} |^3} g^{(n)}(f(z))
\nonumber \\
   &=& \frac{q}{4 \pi \epsilon_1} \frac{d}{|\boldsymbol \rho - d\,\hat{z} |^3}
 \int_{f_2}^{f_1} g^{(n)}(f) d f . 
\eea

The functions $g^{(n)}(f(z))$ up to $n=5$ are

\bea
g^{(1)}(f(z)) &=& 1 
\nonumber \\
g^{(2)}(f(z)) &=& -f(z) + \frac{1}{2}(f_1 + f_2)
\nonumber \\
g^{(3)}(f(z)) &=& \frac{f^2(z)}{2} - \frac{1}{2}(f_1 + f_2)\,f(z) 
                  + \frac{1}{2} f_1 f_2 
\nonumber \\
g^{(4)}(f(z)) &=& -\frac{f^3(z)}{6} + \frac{1}{4}(f_1 + f_2)\,f^2(z) 
                  - \frac{1}{2} f_1 f_2 f(z)
                  - \frac{1}{24}(f_1 +f_2)(f_1^2 - 4 f_1 f_2 + f_2^2)
\nonumber \\
g^{(5)}(f(z)) &=& \frac{f^4(z)}{24} - \frac{1}{12}(f_1 + f_2)\,f^3(z) 
                  + \frac{1}{4} f_1 f_2 f^2(z)
                  + \frac{1}{24}(f_1 +f_2)(f_1^2 - 4 f_1 f_2 + f_2^2) f(z)
\nonumber \\
\label{gns}
 && \hskip 1 in   - \frac{1}{24} f_1 f_2 (f_1^2 - 3 f_1 f_2 + f_2^2).
\eea

We will show by induction that $g^{(n)}(f)$ is

\bea
g^{(n)}(f) &=& (-1)^{n-1} \frac{1}{(n-1)!} f^{n-1} 
      + \frac{1}{2} \left[
      C_1\,g^{(n-1)}(f) - \frac{1}{2!} C_2\,g^{(n-2)}(f) + 
      \frac{1}{3!} C_3\,g^{(n-3)}(f) + \cdots \right. \nonumber \\
 && \hskip 1.5 in \left. + (-1)^{n-2} \frac{1}{(n-1)!} C_{n-1}\,g^{(1)}(f)
                    \right]
\nonumber \\
\label{gn}
           &=& \frac{(-1)^{n-1} f^{n-1}}{(n-1)!}
      + \frac{1}{2} \sum_{m = 1}^{n-1} 
        \frac{(-1)^{n-m-1}C_{n-m}}{(n-m)!} g^{(m)}(f),
\eea
\noindent
where the coefficients $C_n = f_1^n + f_2^n$. First, (\ref{gn}) can be
explicitly verified up to $n=5$  using (\ref{gns}).
Second, we show that if this expression holds for 
some integer $n$, then it also holds for $n+1$. 
From Eq.(\ref{sigman}) we can write,

\bea
g^{(n+1)}(f(z)) &=& \frac{1}{2}
   \left[\int_{f(z)}^{f_1} g^{(n)}(f) d f
       - \int_{f_2}^{f(z)} g^{(n)}(f) d f \right] \nonumber \\
                &=& 
   \frac{(-1)^{n-1}}{(n-1)!} \frac{1}{2}
                \left[\int_{f(z)}^{f_1} f^{n-1} d f
                    - \int_{f_2}^{f(z)} f^{n-1} d f \right] \nonumber \\ 
 && \hskip 0.5 in
  + \frac{1}{2} \sum_{m = 1}^{n-1} 
        \frac{(-1)^{n-m-1} C_{n-m}}{(n-m)!} 
            \frac{1}{2}
       \left[\int_{f(z)}^{f_1} g^{(m)}(f) d f
       - \int_{f_2}^{f(z)} g^{(m)}(f) d f \right] \nonumber \\
                &=& 
        \frac{(-1)^{n} f^{n}}{n!} 
 + \frac{1}{2} \frac{(-1)^{n-1} (f_1^n + f_2^n)}{n!} 
 +      \frac{1}{2} \sum_{m = 1}^{n-1} 
        \frac{(-1)^{n-m-1} C_{n-m}}{(n-m)!} g^{(m+1)}(f) \nonumber \\
                &=& 
\label{gn2}
   \frac{(-1)^{(n+1)-1} f^{(n+1)-1}}{((n+1)-1)!} 
      + \frac{1}{2} \sum_{m = 1}^{(n+1)-1} 
        \frac{(-1)^{(n+1)-m-1} C_{(n+1)-m}}{((n+1)-m)!} g^{(m)}(f)
\eea
\noindent
We thus proved that $g^{(n)}(f)$ is given by (\ref{gn}) for any given integer $n \ge 2$ 
with $g^{(1)}(f)=1$.

We now need to find the integral $\int \sigmai^{(n)}$ in (\ref{sigman}).
We will show by induction that
\be
\int_{f_2}^{f_1} g^{(n)}(f) = -2\frac{E_{n}}{n!}u^{n},
\label{intgn}
\ee
\noindent where $u=f_1-f_2$ and $E_n$ are the coefficients of the expansion
\be
\label{power}
\frac{2}{e^u+1} = \sum_{n=0}^{\infty} \frac{E_n}{n!} u^n.
\ee
\noindent It is easy to see that $E_0=1$.

The base for the mathematical induction for (\ref{intgn}) 
is easily established for the first few terms using (\ref{gns}). 
Now we verify that (\ref{intgn}) holds true for $n+1$ if it is true for $n$.
To do so, we integrate both sides of (\ref{gn2}) and use the
assumption (\ref{intgn}) to obtain
\bea
\int_{f_2}^{f_1} g^{(n+1)}(f) 
&=& -\frac{(-1)^{(n+1)}}{(n+1)!} (f_1^{n+1}-f_2^{n+1}) 
+ \sum_{m = 1}^{n} \frac{(-1)^{n+1-m}}{(n+1-m)! m!} C_{n+1-m} E_m u^{m}
\nonumber \\
&&\hspace{-2cm}= -2\frac{(- f_1)^{n+1}}{(n+1)!} + 
\sum_{m = 0}^{n} \frac{(-f_1)^{n+1-m}}{(n+1-m)!} 
\frac{E_m u^{m}}{m!} +
\sum_{m = 0}^{n} \frac{(-f_2)^{n+1-m}}{(n+1-m)!} 
\frac{E_m u^{m}}{m!} 
\nonumber \\
&&\hspace{-2cm}= -2\frac{(- f_1)^{n+1}}{(n+1)!} + \sum_{m = 0}^{n+1} \left[
 \frac{(-f_1)^{n+1-m}}{(n+1-m)!} + \frac{(-f_2)^{n+1-m}}{(n+1-m)!} \right]
\frac{E_m u^{m}}{m!}  - 2\frac{E_{n+1} u^{n+1}}{(n+1)!}.
\label{int.gn}
\eea
In the second step we have included an $m=0$ term in the summation 
and in the third step we have added and
subtracted an $m=n+1$ term. It can be easily verified that the right
hand side of (\ref{int.gn}) is the $s^{n+1}$ term of the following expression.
\be
-2e^{-f_1 s} + \left[ e^{-f_1 s} + e^{-f_2 s} - 2 \right]\frac{2}{e^{us}+1}
= -2\frac{2}{e^{us}+1} \nonumber \\
= -2\sum_{m=0}^{\infty} \frac{E_m u^m}{m!}s^m.
\ee
\noindent This completes the proof.

Summing over all the terms, we have
\be
\sum_{n=1}^{\infty} \int_{f_2}^{f_1} g^{(n)}(f)
                    = -2\sum_{n=0}^{\infty} \frac{E_n u^n}{n!} + 2 E_0
                    = 2\left(1 - \frac{2}{e^u+1} \right)
= \frac{2(\epsilon_1 - \epsilon_2)}{\epsilon_1 + \epsilon_2}
\ee
We note that the series converges for $|u| = {\rm ln} \epsilono/\epsilon_1 < \pi$. 
This means that if one medium is water ($\epsilono \approx 80$) 
then for the other material the dielectric
constant $\epsilon_1 > \epsilono e^{-\pi} \approx 3.47$. However, using
techniques similar to Borel summation, one can show that 
the series can still be summed to the correct final formula
for larger values of $|u|$.

Finally the induced surface charge density becomes

\be
\sigmai(\boldsymbol \rho) = \frac{q}{4 \pi \epsilon_1}
\frac{2(\epsilon_1 - \epsilon_2)}{\epsilon_1 + \epsilon_2} 
 \frac{d}{|\boldsymbol \rho - d\,\hat{z} |^3}, 
\ee
\noindent which is identical to (\ref{planar.sol}). Thus, we have rigorously
justified using the average dielectric constant $(\epsilon_1 + \epsilon_2)/2$
at the boundary.

\section{Evaluation of $\Lambda$ for Spheres with Sharp Boundaries}

To compute $\Lambda^j_{lm}(a_j,\b L_{j \to k})$, defined as
\be
\Lambda^j_{lm}(a_j,\b L_{j \to k}) \equiv \int {Y_{lm}^*({\tdb r_j -\b L_{j \to k} \over |\tdb r_j -\b L_{j \to k}|}) \over |\tdb r_j -\b L_{j \to k}|^{l+1} }
 \rho_j(\tdb r_j) d \tdb r_j \;,
\ee
for the case of spheres with sharp boundaries,
expand the charge density on sphere $j$ as
\be
\rho_j(\tdb r_j) = \delta(\t r_j - a_j) \sum_{l',m'} \sqrt{4\pi} 
\sigma^j_{l'm'} Y_{l'm'}(\htd r_j)
\ee
to find
\be
\Lambda^j_{lm}(a_j,\b L_{j \to k}) = \sum_{l',m'} \sqrt{4\pi} \sigma^j_{l'm'} 
\int {Y_{lm}^*({\tdb r_j -\b L_{j \to k} \over |\tdb r_j -\b L_{j \to k}|}) 
Y_{l'm'}(\htd r_j) \over 
L_{j \to k}^{l+1} (1+t^2-2t\cos\t\theta_j)^{(l+1)/2} } 
\delta(\t r_j - a_j) d \tdb r_j \; ,
\ee
where use has been made of the geometrical fact that 
$|\tdb r_j -\b L_{j \to k}|=L_{j \to k}\sqrt{1+t^2-2t\cos\t\theta_j}$ with 
$t \equiv \t r_j/L_{j \to k}$.
The delta function renders the radial integration trivial:
\be
\Lambda^j_{lm}(a_j,\b L_{j \to k}) = \sum_{l',m'} \sqrt{4\pi} a_j^2 \sigma^j_{l'm'} 
\int {Y_{lm}^*(\vartheta,\varphi) Y_{l'm'}(\t \theta_j, \t \phi_j)
\over  L_{j \to k}^{l+1} (1+t^2-2t\cos\t\theta_j)^{(l+1)/2}} 
d (\cos \t \theta_j) d \t \phi_j \; ,
\ee
where $\vartheta$ and $\varphi$ are the polar variables of 
$ (\tdb r_j -\b L_{j \to k}) / |\tdb r_j -\b L_{j \to k}|$
and $t=a_j/L_{j \to k}$ now.
All of the angular variables are measured with respect to a coordinate
system whose $z$ axis is parallel to $\b L_{j \to k}$.
The angles $\vartheta$ and $\varphi$ must be expressed as functions of 
the integration variables $\t \theta_j$ and $\t \phi_j$:
\bea
\cos \vartheta & = &
(t \cos \t \theta_j -1) \over \sqrt{1 + t^2 - 2 t \cos \t \theta_j} \\
\varphi & = & \t \phi_j \; .
\eea
Since the definition of the spherical harmonics is
\be
Y_{lm}(\theta, \phi) = \sqrt{\frac{(2l+1)(l-m)!}{4\pi(l+m)!}}
P_{lm}(\cos\theta) e^{im\phi} \; ,
\ee
$\Lambda$ is
\bea
\Lambda^j_{lm}(a_j,\b L_{j \to k}) & = & \sum_{l',m'} 
\frac{\sqrt{4\pi} a_j^2 \sigma^j_{l'm'}}{L_{j \to k}^{l+1}}
\left[ \frac{(2l+1)(l-m)!(2l'+1)(l'-m')!}
{4\pi(l+m)!4\pi(l'+m')!} \right]^{1/2} \\
& \times & \int \frac{P_{lm}
(\frac{(t \cos \t \theta_j -1)}{\sqrt{1 + t^2 - 2 t \cos \t \theta_j}}) P_{l'm'}(\cos \t \theta_j)}
{(1+t^2-2t\cos\t\theta_j)^{(l+1)/2}}
d (\cos \t \theta_j) d \t \phi_j \; .
\eea
The integration over $\t \phi_j$ produces $2\pi\delta_{mm'}$.
The integration over $\cos \t \theta_j$ is then the integral calcuated
by Yu\cite{Yu}.
The final expression for $\Lambda$ is
\be
\Lambda^j_{lm}(a_j,\b L_{j \to k}) = \sum_{l'}
\frac{Q^j_{l'm} t^{l'} (-1)^{l-m}(l+l')! \sqrt{2l+1}}
{L_{j \to k}^{l+1} [4\pi(l+m)!(l'+m)!(l-m)!(l'-m)!(2l'+1)]^{1/2}} \; ,
\ee
where $Q^j_{l'm} \equiv 4 \pi a_j^2 \sigma^j_{l'm}$.


\end{document}